\documentclass{jfm}
\usepackage{graphicx}
\usepackage{epstopdf, epsfig}
\usepackage{soul} 
\usepackage{mathtools}	
\usepackage[usenames, dvipsnames]{color}	

\newcommand{\Vol}{\rotatebox[origin=c]{180}{\ensuremath{A}}}
\DeclarePairedDelimiter\norm{\lVert}{\rVert}%

\shorttitle{Critical impulse for dislodgement}
\shortauthor{S. Maldonado and G.A.M. de Almeida}

\title{Theoretical impulse threshold for particle dislodgement}

\author{Sergio Maldonado\aff{1}
  \corresp{\email{s.maldonado@soton.ac.uk}},
 \and Gustavo A. M. de Almeida\aff{1}}

\affiliation{\aff{1}Faculty of Engineering and Physical Sciences, University of Southampton, Highfield, Southampton SO17 1BJ, U.K. }

\begin{document}

\maketitle

\begin{abstract}
The problem of determining the threshold of motion of a sediment particle resting on the bed of an open channel has historically been dominated by an approach based on the time-space-averaged bed shear stress (i.e. Shields criterion). Recently, experimental studies have promoted an alternative approach to predict the dislodgement threshold, which is based on the impulse of the flow-induced force. Nonetheless, theoretical analyses accompanying these studies result in complex expressions that fail to provide a direct estimate of said impulse threshold. We employ the work-energy principle to derive a prediction of the fundamental impulse threshold that the destabilising hydrodynamic force must overcome in order to achieve full particle dislodgement. For the bed configuration studied, which is composed of spheres, the proposed expression depends on the mobile particle's size and mass, and shows excellent agreement with experimental observations previously published. 
The derivation presented in this paper may thus represent a robust theoretical framework that aids in the re-interpretation of existing data, as well as in the design of future experiments aimed at analysing the importance of hydrodynamic impulse as criterion for particle dislodgement.
\end{abstract}


\section{Introduction} \label{sec:intro}
Accurate estimation of erosion and sediment transport rates in natural, erodible boundaries (e.g. riverine and estuarine beds) is a key and long-standing challenge in Earth surface dynamics and engineering. A basic problem underpinning this challenge is that of determining when a sediment particle resting on the bed will be dislodged by the flow. The classical approach to this problem, based on the well-known work by A. Shields in the 1930's, employs the temporal-spatial average bed shear stress as criterion for particle dislodgement. A review of Shields-based incipient motion studies (see e.g. \citealt{buffington1997} and references therein) illustrates the impossibility of defining an accurate and universal threshold of motion based on time-space averaged shear stress. 
This impossibility stems from the intrinsic complexity of the problem, where turbulent flow typically occurs over a bed surface composed of sediment particles which are inherently heterogeneous in size and shape. 
Thus, recent studies on initiation of motion have shifted towards approaches that investigate the phenomenon at the spatial scale of individual sediment particles (often idealized in shape, such as spheres) and temporal scale of turbulent fluctuations. 
For instance, \cite{kudrolli2016} emphasised the effect of the torque induced by all relevant forces on a spherical particle under steady flow conditions. 
Other studies have focused instead on the influence of turbulence on initiation of particle motion \citep[e.g.][]{heathershaw1985,nelson1995,sumer2003}, attributing the latter to the peak velocity magnitude of hydrodynamic fluctuations. 
Building on this line of thought, \cite{diplas2008} (and subsequent studies discussed below) demonstrated the importance of both magnitude and duration of turbulent fluctuations in determining whether initiation of motion will occur, thus establishing a particle dislodgement criterion based on the impulse of the fluctuating hydrodynamic force (as opposed to its instantaneous maxima). 
The present paper is aimed at providing a theoretical analysis for said impulse-based criterion.

The pioneering experiments by \cite{diplas2008} demonstrated the importance of impulse for particle dislodgement by carefully controlling and systematically manipulating the magnitude and duration of the destabilising force. This work paved the way for other studies supporting the idea of employing impulse (as opposed to either instantaneous or time-averaged forces) as criterion for dislodgement of individual particles. 
Nevertheless, theoretical models derived so far to support this concept \citep{diplas2008,valyrakis2010,valyrakis2013,celik2010} result in complex relations that do not necessarily predict directly the impulse threshold for particle dislodgement for a given bed configuration, thus limiting the interpretation of results obtained from their corresponding, rather insightful experiments. 
For example, both \cite{diplas2008} and \cite{valyrakis2010} analytically studied the problem by considering time-independent hydrodynamic forces acting on a resting particle (in the case of the former, only lift was considered), yielding second-order differential equations of motion and their respective solutions. 
Although these studies relate forces and their durations (not necessarily via simple expressions), they do not provide a direct prediction of the critical impulse for dislodgement; and the condition that the force be time-independent (i.e. a pulse) is unnecessarily restrictive, as we show later. 
\cite{valyrakis2013} and \cite{celik2010} instead invoked energy principles in their respective theoretical analyses. 
\cite{valyrakis2013} employed the concepts of work done on the particle and consequent gain in mechanical energy, but, as with \cite{diplas2008} and \cite{valyrakis2010}, the dislodgement relation obtained does not provide an estimated value for the impulse threshold. 
\cite{celik2010} approached the problem by considering a critical drag force capable of dislodging the particle, which is computed by defining a hypothetical initial velocity that the particle would require in order to gain the necessary kinetic energy to overcome the local elevation threshold for dislodgement. 
This approach is employed to predict a critical initial velocity, with good results, but the assumption of a non-zero initial particle velocity does not represent the condition of initiation of motion from rest; and as with previously discussed studies, a prediction of the critical impulse (not velocity) is not provided (see \S \ref{sec:pseudo}). 
All the above-mentioned approaches also depend critically on empirical coefficients such as drag, lift and energy transfer.

In this paper, we employ the work-energy principle to derive an expression for the magnitude of impulse threshold necessary for particle dislodgement. 
The proposed criterion is defined in terms of the time-varying force exerted on the particle by the flow (and not the flow variables producing said force), which enables us to derive a relation that is simple and independent of empirical coefficients such as drag and lift. The resulting criterion, which depends on the mobile particle's size and mass for the bed setting investigated, shows excellent agreement with previously published experimental observations.

The derivation of the theoretical impulse threshold is detailed in \S \ref{sec:theory}, whereas validation against experiments is presented in \S \ref{sec:validation}. 
Concluding remarks are discussed in \S \ref{sec:conclusions}. 
Complementary derivations are presented in Appendices \ref{appA0} and \ref{appA}.
 
\section{The impulse threshold} \label{sec:theory}

\begin{figure}
	\centering
	\includegraphics[width=0.9\linewidth]{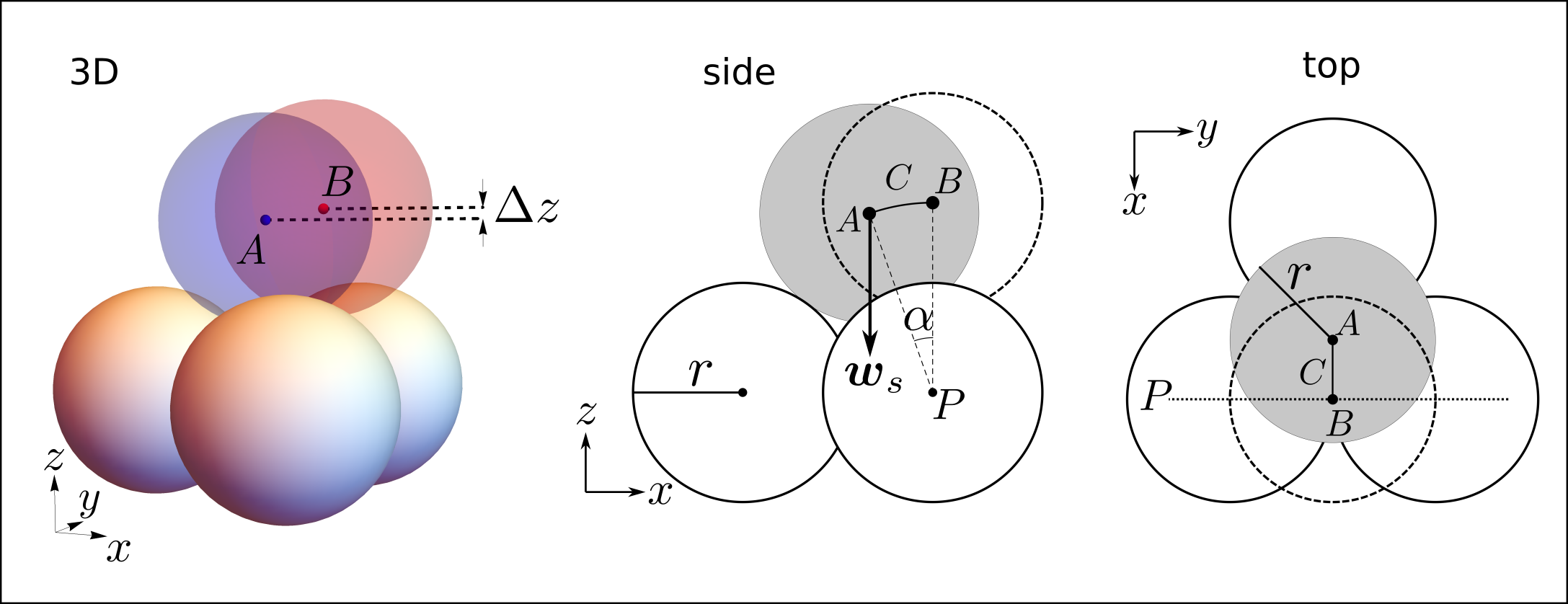}
	\caption{Sketch of the problem. \textit{Left:} 3D rendered image of the setting consisting of four spheres of equal radii $r$, where the mobile particle rests on top of three fixed and densely packed spheres, which in turn lie on a horizontal plane; initial and final (dislodged) positions of the particle's centroid are denoted by points $A$ and $B$, coloured blue and red (online version), respectively. \textit{Middle:} Side view illustrating the trajectory $C$ followed by the particle during dislodgement, and the centroid's angular displacement $\alpha$ with respect to the pivot axis $P$. \textit{Right:} Top view.}
	\label{fig:sketch}
\end{figure}

We are concerned with the conditions leading to full particle dislodgement, defined as the event when a sediment particle originally at rest on the bed surface is transported to a different location on the bed (hereinafter we avoid the alternative terms `incipient motion' or `initiation of motion', which may evoke local particle movement not necessarily leading to a different resting position). 
We focus on the bed configuration employed in reference works \citep{diplas2008,valyrakis2010,valyrakis2013,celik2010,celik2013}, where a mobile spherical particle rests on top of three, equal-sized, fixed, well-packed spheres, as depicted in fig. \ref{fig:sketch} (the same diameter of both top and base spheres is assumed unless otherwise stated). A Cartesian frame of reference is adopted, where $x,y$ and $z$ denote streamwise, transverse and vertical coordinates, respectively. 
Fundamentally, to achieve full dislodgement, the work done on the particle by the net external forces must be sufficient to overcome the elevation threshold resulting from the local micro-topography. This happens when the particle originally at rest in stable equilibrium (point $A$) moves to a higher, unstable position, where an infinitesimally small streamwise force acting at its centre of mass will lead to a new resting position (point $B$ -- a \textit{separatrix} in the context of oscillators). 
Therefore, the minimum work required for full dislodgement can be defined for the condition of the sphere reaching point $B$ with null kinetic energy. 
It is assumed that the particle is dislodged by rolling from $A$ to $B$, which is in agreement with experimental evidence that highlights this entrainment mode as the most common for near-threshold conditions \cite[see e.g.][]{fenton1977,celik2010,kudrolli2016}. 
Moreover, it may be argued that this trajectory and type of motion (as opposed to sliding) minimises the energy required to get to $B$ from $A$, and is thus in line with our objective of finding a fundamental energy threshold for dislodgement. 
Since the sphere rolls without sliding, static friction forces act at the points of contact between the mobile and the base spheres, which in turn create a resultant torque about the rolling particle's centre of mass. However, since the mobile sphere arrives at point $B$ with no angular velocity (we assume null kinetic energy at point $B$), no net work is done by this torque on the sphere, and so we exclude static friction forces from our derivation below. 
For partly-exposed particles like the one under consideration, the net hydrodynamic force may not act exactly at the particle's centre of mass. However, without much knowledge on the actual line of action of said force \citep{kudrolli2016}, we assume that it does for simplicity, as typically done in similar studies \cite[e.g.][]{kudrolli2016, valyrakis2010,celik2010}.

Consider the net force acting on the mobile sphere's centre of mass at any time $t$ during a dislodgement event of duration $T = t_1 - t_0$, $\boldsymbol{F}(t) = \boldsymbol{F}_H(t) +  \boldsymbol{w}_s + \sum \boldsymbol{N}_i(t)$; where $\boldsymbol{F}_H(t)$ is the net hydrodynamic force (see discussions below); $\boldsymbol{w}_s = (\rho_s - \rho) \Vol \boldsymbol{g}$ is the submerged weight of the particle of volume $\Vol$ and density $\rho_s$, immersed in a fluid of density $\rho$, with $\boldsymbol{g}$ representing the gravitational acceleration; and $\sum \boldsymbol{N}_i(t)$, with $i=1,2,...,p$, is the sum of $p$ normal reaction (constraint) forces acting at the $p$ contact points between the mobile and base particles (e.g. in fig. \ref{fig:sketch}, $p=2$ for $t_0 < t \le t_1$). 
Applying the work-energy principle to the sphere moving along the trajectory $C=\boldsymbol{s}(t)$ connecting points $A$ and $B$, and recalling that its change in kinetic energy is taken as null, yields $\int_{C} \boldsymbol{F}(t) \boldsymbol{\cdot} d\boldsymbol{s} = \int_{C} \boldsymbol{F}_H(t) \boldsymbol{\cdot} d\boldsymbol{s} + \int_{C} \boldsymbol{w}_s \boldsymbol{\cdot} d\boldsymbol{s} + \sum \int_{C} \boldsymbol{N}_i(t) \boldsymbol{\cdot} d\boldsymbol{s}= 0 $; which, after noting that $\boldsymbol{N}_i \boldsymbol{\cdot} d\boldsymbol{s} = 0$, leads to the dislodgement condition
\begin{equation} \label{eq:1}
\int_{C} \boldsymbol{F}_H(t) \boldsymbol{\cdot} d\boldsymbol{s} = \int_{t_0}^{t_1} \boldsymbol{F}_H(t) \boldsymbol{\cdot} \frac{d \boldsymbol{s}}{d t} dt \, > w_s \, \Delta z,
\end{equation}
where $w_s \equiv \norm{\boldsymbol{w}_s }$ and $\Delta z$ is the elevation gained by the particle's centre of mass (i.e. the vertical distance between points $A$ and $B$). 
Since the particle is at rest at $t_0$, its velocity $\boldsymbol{v}(t) \equiv d \boldsymbol{s}(t) / dt$ relates, through Newton's second law, to its own mass and acceleration, $m$ and $\boldsymbol{a}(t)$, respectively, and $\boldsymbol{F}(t)$ as follows:
\begin{equation} \label{eq:vel}
\boldsymbol{v}(t) \equiv \int_{t_0}^{t} \boldsymbol{a}(\tau) d \tau = \frac{1}{m} \int_{t_0}^{t} \boldsymbol{F}(\tau) d \tau .
\end{equation}

Substitution of \eqref{eq:vel} in \eqref{eq:1} yields the following dislodgement criterion in terms of time integrals of all relevant forces (i.e. impulses):
\begin{equation} \label{eq:disfull}
\int_{t_0}^{t_1} \boldsymbol{F}_H(t) \boldsymbol{\cdot} \left [ \int_{t_0}^{t} \boldsymbol{F}_H(\tau) d \tau + \boldsymbol{w}_s (t-t_0) + \sum_i \int_{t_0}^{t} \boldsymbol{N}_i(\tau) d \tau  \right ] dt >  m w_s \Delta z .
\end{equation}

Eq. \eqref{eq:disfull} represents an exact condition to be verified if dislodgement is to take place. However, in this form it is of no practical use. The above relation can be simplified under the assumption that the mobile sphere is highly exposed to the flow, such that i) the angular displacement of the particle's centroid during dislodgement with respect to the pivot axis $P$ is small (see $\alpha$ in fig. \ref{fig:sketch}); and ii) $\boldsymbol{F}_H$ is dominated by the drag component acting predominantly in the streamwise ($x$) direction, which is supported by experimental evidence on similar bed configurations \citep{fenton1977,celik2010,schmeeckle2007}. 
Under these assumptions, approximately $\boldsymbol{F}_H \perp \boldsymbol{w}_s$ and $\boldsymbol{F}_H \perp \boldsymbol{N}_i$ for all $t$ during the dislodgement event, thus leading to vanishing of the second and third dot products in the left-hand-side of \eqref{eq:disfull}. (See Appendix \ref{appA0} for a more rigorous treatment of eq. \ref{eq:disfull} and discussion on these assumptions.) 
Further noting that $\boldsymbol{F}_H(t) \boldsymbol{\cdot} \boldsymbol{F}_H(\tau)$ represents a symmetric function $f(t,\tau)$, such that $f(t,\tau) = f(\tau,t)$, we obtain

\begin{equation} \label{eq:intlimits}
\int_{t_0}^{t_1} \boldsymbol{F}_H(t) \, \boldsymbol{\cdot} \int_{t_0}^{t} \boldsymbol{F}_H(\tau) d \tau dt = \frac{1}{2} \int_{t_0}^{t_1} \int_{t_0}^{t_1} f \; d \tau dt = \frac{1}{2} \int_{t_0}^{t_1} \boldsymbol{F}_H(t) dt \, \boldsymbol{\cdot} \int_{t_0}^{t_1} \boldsymbol{F}_H(\tau) d \tau .
\end{equation}

Combination of \eqref{eq:intlimits} and the assumption of approximate orthogonality between $\boldsymbol{F}_H$ and both $\boldsymbol{w}_s$ and $\boldsymbol{N}_i $ discussed above, permits significant simplification of \eqref{eq:disfull} (see Appendix \ref{appA0}). 
This simplification represents an approximate estimate of the magnitude of the critical impulse imparted to the particle during $T$, $J_c$, that a destabilising hydrodynamic force (capable of doing work on the particle) must overcome in order to achieve full dislodgement; namely:
\begin{equation} \label{eq:Ic1}
\norm[\bigg]{ \int_{t_0}^{t_1} \boldsymbol{F}_H(t) dt } \equiv J_c \approx \sqrt{2 m w_s \Delta z} \; .
\end{equation}

The elevation threshold $\Delta z$ is left as a free parameter in \eqref{eq:Ic1} for reasons that become clear in \S \ref{sec:pseudo}. 
Note, however, that for certain ideal configurations, $\Delta z$ has analytical solutions. 
For instance, data employed for validation in \S \ref{sec:direct} are derived from the experimental setting depicted in fig. \ref{fig:sketch}, for which it can be shown that (see Appendix \ref{appA})
\begin{equation} \label{eq:dz}
\Delta z = \left ( \frac{3 - 2 \sqrt{2}}{\sqrt{3}} \right ) r ,
\end{equation}
where $r$ is the radius of the spheres. 
The mass of the mobile sphere, $m$, appearing in \eqref{eq:Ic1} merits some discussion. If $\boldsymbol{F}_H (t)$ is taken as the total hydrodynamic force exerted on the mobile particle by the surrounding fluid (that is, excluding the buoyancy force, which is already accounted for in $\boldsymbol{w}_s$) -- i.e. the integral of pressure and shear stresses over the surface of the sphere (minus the buoyancy force) -- then $m$ represents the actual mass of the sphere, i.e. $m = \rho_s \Vol$. 
However, direct measurement of $\boldsymbol{F}_H (t)$, thus defined, is extremely challenging. For this reason, experimentalists and modellers alike typically estimate this force via parametrisations strictly applicable to non-accelerating particles (drag and lift coefficients being the prime examples). 
This in turn prompts the need to account for the effect of the particle's acceleration separately, which is typically done via the added mass coefficient, $M$, by modifying the sphere's mass as $m=(\rho_s + M \rho) \Vol$. The theoretical value of $M=0.5$ for a sphere, arising from potential flow theory, is commonly employed in studies dealing with the dynamics of bed particles (see e.g. \citealt{barati2018} for a recent review on saltating particle models). Although derived from inviscid flow theory, the value of $M \approx 0.5$ also appears to be supported by experiments with spheres in real (viscous) flows \cite[see e.g.][]{pantaleone2011}. In general, if the added mass effect must be considered (either through a theoretically or empirically defined value of $M$) due to the treatment of $\boldsymbol{F}_H$ discussed above, \eqref{eq:Ic1} becomes
\begin{equation} \label{eq:Ic2b}
{J}_c \approx \frac{4}{3} \rho \pi r^3 \;  \sqrt{2 \left ( s+M \right )\left ( s-1 \right ) g \Delta z} \; ,
\end{equation}
with $s \equiv \rho_s / \rho$ being the particle's relative density and $g \equiv \norm{\boldsymbol{g}} $. 
The proposed criterion for critical impulse, derived for spherical particles, is solely dependent on the particle's size and density (that is, if $M=0.5$ is adopted), and is applicable to a hydrodynamic time-varying force of arbitrary duration. 
This contrasts with expressions previously derived for the same bed configuration, as discussed in \S \ref{sec:intro}. 
The (relative) lack of empirical coefficients in our derivation is in good measure achieved by the very definition of the criterion, which expresses the dislodgement condition as a function of the hydrodynamic force exerted on the particle, rather than the flow variables inducing said force. 
Naturally, coefficients such as drag will be necessary in practice when relating the fluid force exerted on the particle to local hydrodynamic parameters (see \S \ref{sec:pseudo}). 
However, this is beyond the scope of the present paper, the main aim of which is to provide a theoretical analysis of existing experiments devoted to exploring the role of hydrodynamic impulse as criterion for dislodgement.

Even though derivation of \eqref{eq:Ic1} implicitly assumes a horizontal channel (the simplifying assumptions invoked depend on this condition), the effect of the local slope is accounted for via $\Delta z$, which will be affected by variations in the local micro-topography. 
However, use of  \eqref{eq:Ic1} and \eqref{eq:Ic2b} should be restricted to applications involving horizontal or nearly-horizontal channels, given that terms neglected in \eqref{eq:disfull} are anticipated to grow in importance for steep slopes. 
In deriving \eqref{eq:Ic1}, we have also assumed exclusively forces capable of doing work on the particle (but note that this is not a restriction of the full condition, eq. \ref{eq:disfull} -- i.e. hydrodynamic forces too small to move the particle will simply not comply with the inequality). 
For the setting considered (fig. \ref{fig:sketch}), the maximum force opposing motion will be found at the initial position (see end of \S \ref{sec:FvsT}).

\section{Comparison against experiments} \label{sec:validation}

In order to test the validity of the theoretical impulse threshold for particle dislodgement proposed here, we compare predictions of \eqref{eq:Ic1} and \eqref{eq:Ic2b} against experiments by \cite{celik2013}, \cite{celik2010}, \cite{diplas2008} and \cite{valyrakis2010}. 
All these experiments deal with a setting similar to that illustrated in fig. \ref{fig:sketch}, where a mobile spherical particle rests on top of three, fixed, well-packed spheres. 
\cite{celik2010,celik2013} employed identical-size top and base spheres, with $s=2.3$ and $r = 6.35$ mm, subject to water flow. 
On the other hand, \cite{diplas2008} and \cite{valyrakis2010} investigated different combinations of top and base metallic spheres' sizes subject to electromagnetic flux. 
For each of the above experiments, the parameters measured and methodologies vary, allowing us to test the derived theoretical prediction of critical impulse under different conditions, as detailed next.

\subsection{Direct comparison (impulse)} \label{sec:direct}
The most direct comparison is made against the experiments by \cite{celik2013}, who measured pressure time histories at four points on the surface of a stationary sphere, which were then used to obtain a direct estimate of the force (and thus impulse) responsible for the dislodgement of another, otherwise-identical mobile sphere subject to the same flow conditions. 
\cite{celik2013} estimate impulse threshold from measurements of the (streamwise-aligned) drag force, in line with the assumptions we have invoked in \S \ref{sec:theory}. 
We select for comparison run U8 in the referred publication, which reports the lowest frequency of particle dislodgement events observed (namely, 0.14 dislodgement events per minute), which can be interpreted as the experimental conditions that are closest to a fundamental threshold for dislodgement. 
For run U8, \cite{celik2013} report a mean value of the critical impulse of 0.0002 Ns. 
For this experimental setting, use of \eqref{eq:Ic2b}, with $M=0.5$ and $\Delta z$ computed from \eqref{eq:dz}, yields a prediction of the critical impulse of $2.27 \times 10^{-4}$ Ns. In other words, the proposed criterion shows a virtually exact agreement (to the precision reported by \citealt{celik2013}) with these experimental results.

\subsection{Indirect comparison (pseudo-impulse)} \label{sec:pseudo}
Next, we consider the experiments by \cite{celik2010}, who measured the streamwise component of the fluid velocity, $u$, one particle diameter upstream of a test (mobile) sphere. 
\cite{celik2010} studied particle dislodgement as a function of pseudo-impulse, defined as the product $\left\langle u^2 \right\rangle  T$, where the angle brackets denote time-averaging over the interval $T$. 
The referred researchers employ this surrogate for impulse based on the proviso that the prevailing hydrodynamic force is drag, $F_D$, which is proportional to $u^2$. 
Considering the time-average of the net hydrodynamic force acting on the particle over the interval $T$ ,
\begin{equation} \label{eq:timeav}
\left \langle \boldsymbol{F}_H \right \rangle \equiv \frac{1}{T} \int_{t_0}^{t_1} \boldsymbol{F}_H(t) dt ,
\end{equation}
the impulse imparted to the particle by the flow over $T$ can also be expressed as $\boldsymbol{J} =  \left \langle \boldsymbol{F}_H \right \rangle T $. 
We can then find an approximate equivalence between the real impulse, $\boldsymbol{J}$, and pseudo-impulse employed by \cite{celik2010}, $\boldsymbol{J}_{ps}$, as follows:
\begin{equation} \label{eq:pseudo}
\norm{\boldsymbol{J}_{ps}} \equiv \left \langle u^2 \right \rangle T = \frac{ \left \langle F_D \right \rangle T}{ \frac{1}{2} \rho A_p C_D} \approx \frac{ \norm{ \boldsymbol{J}}}{ \frac{1}{2} \rho A_p C_D} ,
\end{equation}
where the conventional parametrisation of the drag force has been employed; i.e. $F_D = 0.5 \rho A_p C_D u^2$, with $C_D$ representing the drag coefficient and $A_p$ being the projected area of the spherical particle. 
The latter may be approximated, due to the assumption of the particle being highly exposed to the flow (also employed by \citealt{celik2010}), as $A_p = \pi r^2$. 
The assumption of drag being the predominant hydrodynamic force acting on the particle underpins \eqref{eq:pseudo}. 

In their experiments, \cite{celik2010} observe a range of values of pseudo-impulse of 0.0034 to 0.0095 m$^2$/s, for which both dislodgement and no-dislodgement events are observed for all values of $\left \langle u^2 \right \rangle$ considered. 
In other words, below(above) this range of pseudo-impulse, the test particle was never(always) dislodged by the flow. 
Therefore, the lower limit of this range (0.0034 m$^2$/s) can be interpreted as the fundamental pseudo-impulse threshold, below which no particle dislodgement is observed. 

To test our prediction of critical impulse \eqref{eq:Ic2b}, we first convert it to pseudo-impulse via \eqref{eq:pseudo}, and use $\Delta z = 0.4$ mm. 
The reason for fixing $\Delta z$ to the value reported by \cite{celik2010} is that their experimental setting included the presence of a retention pin downstream of the mobile particle that ensured that once `fully dislodged', the particle could return to its original position, thus automating the experiment. 
To transform impulse to pseudo-impulse, a value of $C_D$ must be assumed (see eq. \ref{eq:pseudo}). 
To this end, we employ the value of $C_D$ from run U8 in \cite{celik2013} (namely, $C_D = 0.818$), who carried out very similar experiments to those of \cite{celik2010}, reporting values of $C_D$ for a resting sphere under diverse flow conditions. 
As in \S \ref{sec:direct}, our focus is on run U8 because this run represents the experimental conditions which are closest to a fundamental dislodgement threshold. 
Use of \eqref{eq:Ic2b} in \eqref{eq:pseudo}, with $C_D = 0.818$, $M=0.5$ and $\Delta z = 0.4$ mm, yields a prediction of critical pseudo-impulse of 0.0035 m$^2$/s, which shows a very good agreement with the lower limit experimentally found by \cite{celik2010} (i.e. 0.0034 m$^2$/s), thus supporting the argument that \eqref{eq:Ic2b} describes accurately the fundamental impulse threshold for particle dislodgement.

\cite{celik2010} also proposed an algorithm to predict the critical pseudo-impulse, which yielded an estimate of 0.0033 m$^2$/s for this experimental setting. 
This estimate is very close to our prediction of 0.0035 m$^2$/s, but it is important to highlight that the method proposed by \cite{celik2010} i) represents a methodology, rather than an expression, to estimate the critical pseudo-impulse; ii) lacks rigour by requiring a hypothetical \textit{initial velocity} of the \textit{resting} particle; and iii) needs input of certain geometric variables (such as lever arms) not required in \eqref{eq:Ic2b}. 

\subsection{Recovery of trend $F$ vs $T$ obtained empirically} \label{sec:FvsT}
\cite{diplas2008} were the first to demonstrate experimentally the importance of force duration by plotting normalized drag force, $\hat{F}_D$, versus its normalized duration, $\hat{T}_D$. 
A metallic particle subject to electromagnetic flux was employed to achieve a highly-controllable flow. 
Measurements by \cite{diplas2008} (328 in total) collapsed remarkably well into a curve of the form $\hat{F}_D = K {\hat{T}_D}^n$, where $K$ and $n$ are constants obtained from best-fit curves, the latter of which takes a value of $n \approx -1$ ($n=-0.99$, to be precise). 
Later, \cite{valyrakis2010} extended this study and provided different values of coefficients $K$ and $n$ arising from best-fit curves for different combinations of top and base spheres' diameters (a total of 1709 data points was obtained); the value of $n \approx -1$ was confirmed: their reported values of $n$ range from $-0.89$ to $-1.07$ with a mean of $-0.99$ 
(\citealt{valyrakis2010} employ squared voltage across the electromagnet as proxy for force). 
We conclude our experimental comparisons by noting that the value of $n=-1$ is to be expected since, by rewriting \eqref{eq:Ic1} as $J_c = \norm{\left \langle \boldsymbol{F}_H \right \rangle} T \approx \sqrt{2 m w_s \Delta z}$, we obtain
\begin{equation} \label{eq:FvsT}
\norm{\left \langle \boldsymbol{F}_H \right \rangle} \approx \left ( \sqrt{2 m w_s \Delta z} \right ) \frac{1}{T} ,
\end{equation}
where $\sqrt{2 m w_s \Delta z}$ is indeed a constant for a local bed configuration (i.e. $K$) and $n$ is precisely equal to $-1$. 
Combinations of $\norm{\left \langle \boldsymbol{F}_H \right \rangle}$ and  $T$ falling above the curve given by \eqref{eq:FvsT} will result in particle dislodgement, so long as the net hydrodynamic force is capable of doing work on the particle during $T$. 
For the problem considered, where $\boldsymbol{F}_H$ equals a virtually time-independent constant $\widetilde{F_H}$ over $T$ (a pulse), the condition to verify is $\widetilde{F_H} \cos \alpha_0 > w_s \sin \alpha_0$, where $\alpha_0 = \alpha (t=t_0)$; or $\widetilde{F_H} \gtrapprox w_s / 3$ for the case of equal-sized top and base spheres (in fig. \ref{fig:sketch}, the maximum force opposing motion is found at  $t_0$, where $\sin \alpha_0 = 1/3$). 
It is worth remarking that \eqref{eq:FvsT} is only approximate. In Appendix \ref{appA0}, an exact expression is derived for this experimental setting (eq. \ref{eq:dislmagn}), which is completely determined if the time history of $\boldsymbol{a}(t)$ (or the particle's position) is known.  
However, as discussed in the same appendix, said exact equation reduces to the approximation proposed here (eq. \ref{eq:Ic1}) for small angular displacement of the particle during dislodgement, as is assumed to be the case with the highly-exposed sphere under consideration. 
A quantitative comparison between the approximate constant predicted (i.e. $\sqrt{2 m w_s \Delta z}$) and $K$ empirically found by \cite{valyrakis2010} cannot be carried out without more detailed information on the experiment, especially pertaining to the electromagnet (e.g. resistance, number of turns in the coil, etc.), which unfortunately is unavailable.

\section{Conclusions} \label{sec:conclusions}
We propose a theoretical estimate of the critical impulse of the destabilising hydrodynamic force that must be exceeded to achieve particle dislodgement. 
The proposed expression, derived from the work-energy principle and valid for the bed setting depicted in fig. \ref{fig:sketch} and assumptions discussed in \S \ref{sec:theory}, represents a scalar value that depends exclusively on the local bed arrangement. 
The derived impulse-based criterion for dislodgement shows excellent agreement with previously published experimental data by: i) yielding a virtually exact prediction of the critical impulse reported by \cite{celik2013}; ii) predicting well the fundamental threshold for particle dislodgement, even when converted to pseudo-impulse (as defined in \S \ref{sec:pseudo}), experimentally observed by \cite{celik2010}; and iii) naturally recovering the trend $F \propto T^{-1}$ obtained approximately via best-fit curves by \cite{diplas2008} and \cite{valyrakis2010} after analysing a total of 2037 data points combined. 
The remarkable agreement between the theory here derived and experimental data is encouraging, especially in view of the notorious uncertainty associated with the prediction of initiation of sediment motion. However, availability of relevant empirical data is still rather limited. 
We hope, therefore, that the present work may be used as theoretical framework that aids in the design of future experiments aimed at continuing the investigation on the importance of hydrodynamic impulse as criterion for particle dislodgement, which will in turn help testing the present theory further.
\\\\
Acknowledgements: We would like to thank the contribution of the editor and two anonymous referees, whose insightful reviews of the paper have helped improve the overall quality of the manuscript. SM also wishes to thank Dr. Joanna A. Zieli\'{n}ska for her helpful advice on some of the derivations.

\appendix
\section{}\label{appA0}
We explore here the simplifications adopted to arrive at \eqref{eq:Ic1} from \eqref{eq:disfull}. Referring to the setting depicted in fig. \ref{fig:sketch}, motion takes place in the $x-z$ plane, such that for $t>t_0$, $\sum \boldsymbol{N}_i (t) \equiv \boldsymbol{N}(t) = N_x(t) \boldsymbol{\hat{i}} +  0 \boldsymbol{\hat{j}} + N_z(t) \boldsymbol{\hat{k}}$, where $\boldsymbol{\hat{i}}$, $\boldsymbol{\hat{j}}$ and $\boldsymbol{\hat{k}}$ are the unit vectors pointing in the $x$, $y$ and $z$ directions, respectively. 
In general, we have $\boldsymbol{F}_H (t) = F_{Hx}(t) \boldsymbol{\hat{i}} +  F_{Hy}(t) \boldsymbol{\hat{j}} + F_{Hz}(t) \boldsymbol{\hat{k}}$, $\boldsymbol{a} (t) = a_{x}(t) \boldsymbol{\hat{i}} +  a_{y}(t) \boldsymbol{\hat{j}} + a_{z}(t) \boldsymbol{\hat{k}}$ and $\boldsymbol{w}_s = -w_s \boldsymbol{\hat{k}}$. 
At any time, the position of the particle is completely determined by the angle $\alpha (t)$ formed by the straight line from the mobile particle's centroid to the pivot axis $P$ and the vertical (see fig. \ref{fig:sketch}), which varies from $\alpha(t=t_0) = \alpha_0$ to $\alpha(t=t_1)=\alpha_1=0$. 
For convenience let us define, for any function $g(\theta)$, the operator 
\begin{equation}
\left \langle g(\theta) \right \rangle_\omega \equiv \int_{t_0}^{\omega} g(\theta) d \theta ,
\end{equation}
which allows us to write the left-hand-side of \eqref{eq:disfull} as 
\begin{equation}
\left \langle \boldsymbol{F}_H(t) \boldsymbol{\cdot} \left [  \left \langle \boldsymbol{F}_H(\tau) \right \rangle_t +  \left \langle \boldsymbol{w}_s \right \rangle_t  +  \left \langle \boldsymbol{N}(\tau) \right \rangle_t \, \right ] \; \right \rangle_{t_1} .
\end{equation}

As discussed in \S \ref{sec:theory}, the first dot product can be written as $0.5 \left \langle F_H \right \rangle^2_{t_1}$ (where $F_H \equiv \norm{\boldsymbol{F}_H}$), so our focus here is on the second and third dot products. 
Defining $\boldsymbol{\hat{n}}(t) = n_x(t) \boldsymbol{\hat{i}} + n_z(t) \boldsymbol{\hat{k}}$ as the unit vector pointing in the direction of $\boldsymbol{N}(t)$, the sum of forces along $\boldsymbol{\hat{n}}(t)$ yields $ \boldsymbol{N}(t) = [(m \boldsymbol{a}(t) -  \boldsymbol{w}_s - \boldsymbol{F}_H(t)) \boldsymbol{\cdot} \boldsymbol{\hat{n}}(t)] \boldsymbol{\hat{n}}(t)$. 
We can then write the products of interest as follows:
\begin{equation} \label{eq:2dot}
\left \langle \boldsymbol{F}_H(t) \boldsymbol{\cdot}    \left \langle \boldsymbol{w}_s \right \rangle_t   \right \rangle_{t_1} = - w_s \left \langle F_{Hz}(t)\, [t-t_0] \right \rangle_{t_1} 
\end{equation}
and (omitting functions' arguments for clarity)
\begin{eqnarray} \label{eq:3dot}
\left \langle \boldsymbol{F}_H \boldsymbol{\cdot}  \left \langle \boldsymbol{N} \right \rangle_t  \, \right \rangle_{t_1} &=& \left \langle \boldsymbol{F}_H \boldsymbol{\cdot}  \left \langle (m \boldsymbol{a} \boldsymbol{\cdot} \boldsymbol{\hat{n}}) \boldsymbol{\hat{n}} \right \rangle_t  \, \right \rangle_{t_1} + \nonumber \\
& & \left \langle \boldsymbol{F}_H \boldsymbol{\cdot}  \left \langle -(\boldsymbol{w}_s \boldsymbol{\cdot} \boldsymbol{\hat{n}}) \boldsymbol{\hat{n}} \right \rangle_t  \, \right \rangle_{t_1} + \left \langle \boldsymbol{F}_H \boldsymbol{\cdot}  \left \langle -(\boldsymbol{F}_H \boldsymbol{\cdot} \boldsymbol{\hat{n}}) \boldsymbol{\hat{n}} \right \rangle_t  \, \right \rangle_{t_1} \nonumber \\
&=& m \left \langle F_{Hx} \left \langle a_x n^2_x + a_z n_x n_z  \right \rangle_t + F_{Hz} \left \langle a_x n_x n_z + a_z n^2_z \, \right \rangle_t \, \right \rangle_{t_1} + \nonumber \\
& & w_s \left \langle F_{Hx} \left \langle n_x n_z \right \rangle_t + F_{Hz} \left \langle n^2_z \right \rangle_t \; \right \rangle_{t_1} - \nonumber \\
& & \left \langle F_{Hx} \left \langle F_{Hx} n^2_x + F_{Hz} n_x n_z  \right \rangle_t + F_{Hz} \left \langle F_{Hx} n_x n_z + F_{Hz} n^2_z \, \right \rangle_t \, \right \rangle_{t_1} 
\end{eqnarray}

The assumption invoked in \S \ref{sec:theory} that $\boldsymbol{F}_H$ acts predominantly in the $x$-direction, such that $\boldsymbol{F}_H \approx F_{Hx} \boldsymbol{\hat{i}}$ (or $F_{Hz} \approx 0$), leads to vanishing of \eqref{eq:2dot} and reduces \eqref{eq:3dot} to $\left \langle m  F_{Hx} \left \langle a_x n^2_x + a_z n_x n_z  \right \rangle_t + w_s F_{Hx} \left \langle n_x n_z \right \rangle_t - F_{Hx} \left \langle F_{Hx} n^2_x \right \rangle_t \right \rangle_{t_1}$. 
The second proviso discussed in \S \ref{sec:theory} states that the angular displacement $(\alpha_1 - \alpha_0)$ is small, such that $n_x \to 0$ (or $n_z \to 1$), thus further leading to $n_x n_z  \to 0$ and $n^2_x \to 0$, and hence vanishing of \eqref{eq:3dot} altogether.

The above discussion justifies simplifying \eqref{eq:disfull} to \eqref{eq:Ic1} (for the conditions stated in \S \ref{sec:theory}) when no detailed information (time series) of all relevant forces is available, which is the anticipated case for most experimental studies. 
However, for highly-controlled experiments such as those discussed in \S \ref{sec:FvsT}, where $\boldsymbol{F}_H$ has a time-independent magnitude $\widetilde{F_H}$ and constant direction $\boldsymbol{\hat{i}}$, the treatment of the full dislodgement condition presented in this appendix may be useful. 
Under said conditions, \eqref{eq:disfull} becomes
\begin{equation} \label{eq:dislmagn}
\widetilde{F_H} \left [ \frac{\widetilde{F_H}}{2} T^2 + \left \langle m   \left \langle a_x n^2_x + a_z n_x n_z  \right \rangle_t    +   w_s  \left \langle n_x n_z \right \rangle_t - \widetilde{F_H} \left \langle n^2_x \right \rangle_t \right \rangle_{t_1} \right ] > m w_s \Delta z ,
\end{equation}

which, naturally, reduces to \eqref{eq:Ic1} if terms $n_x n_z$ and $n^2_x$ are neglected as before.

\section{}\label{appA}
Consider fig. \ref{fig:sketch}, where all spheres have equal radii $r$. 
Connecting the centroids of all four spheres at the initial position draws a regular tetrahedron with edge length $2r$. 
The distance from any vertex of the base (which is an equilateral triangle) to its own centroid, $c$, is thus $2 r / \sqrt{3}$. 
Then, the vertical distance from $c$ to the centroid of the upper sphere (point $A$) is $2 \sqrt{2} r / \sqrt{3}$. 
At the dislodged position, it is readily seen that the vertical distance from the plane where the centroids of the base spheres lie to the centroid of the upper (dislodged) sphere (point $B$) is $\sqrt{4 r^2 - r^2} = \sqrt{3} r$. 
The change in vertical distance from the initial to the dislodged position, $\Delta z$, is therefore $\sqrt{3} r - 2 \sqrt{2} r / \sqrt{3} = (3 - 2 \sqrt{2}) r / \sqrt{3}$ (eq. \ref{eq:dz}). 
Similarly, if we consider spheres of different size, such that the top(base) sphere(spheres) has(have) a radius $r_2$($r_1$), where $r_2 > r_1$, the same procedure yields $\Delta z = \sqrt[]{2 r_1 r_2 + r^2_2} - \sqrt[]{2 r_1 r_2 + r^2_2 - r^2_1 / 3}$ . 

\bibliographystyle{jfm}
\bibliography{impulse}

\begin{thebibliography}{14}
\expandafter\ifx\csname natexlab\endcsname\relax\def\natexlab#1{#1}\fi
\def\au#1{#1} \def\ed#1{#1} \def\yr#1{#1}\def\at#1{#1}\def\jt#1{\textit{#1}}
  \def\bt#1{#1}\def\bvol#1{\textbf{#1}} \def\vol#1{#1} \def\pg#1{#1}
  \def\publ#1{#1}\def\arxiv#1{#1}\def\org#1{#1}\def\st#1{\textit{#1}}

\bibitem[Barati {\em et~al.\/}(2018)Barati, Neyshabouri \& Ahmadi]{barati2018}
{\sc \au{Barati, Reza}, \au{Neyshabouri, Seyed Ali Akbar~Salehi} \& \au{Ahmadi,
  Goodarz}} \yr{2018}  \at{Issues in eulerian--lagrangian modeling of sediment
  transport under saltation regime}.  \jt{International Journal of Sediment
  Research} .

\bibitem[Buffington \& Montgomery(1997)]{buffington1997}
{\sc \au{Buffington, John~M} \& \au{Montgomery, David~R}} \yr{1997}  \at{A
  systematic analysis of eight decades of incipient motion studies, with
  special reference to gravel-bedded rivers}.  \jt{Water Resources Research}
  \bvol{33}~(8),  \pg{1993--2029}.

\bibitem[Celik {\em et~al.\/}(2013)Celik, Diplas \& Dancey]{celik2013}
{\sc \au{Celik, Ahmet~Ozan}, \au{Diplas, Panayiotis} \& \au{Dancey, Clint~L}}
  \yr{2013}  \at{Instantaneous turbulent forces and impulse on a rough bed:
  Implications for initiation of bed material movement}.  \jt{Water Resources
  Research}  \bvol{49}~(4),  \pg{2213--2227}.

\bibitem[Celik {\em et~al.\/}(2010)Celik, Diplas, Dancey \&
  Valyrakis]{celik2010}
{\sc \au{Celik, Ahmet~O}, \au{Diplas, Panayiotis}, \au{Dancey, Clinton~L} \&
  \au{Valyrakis, Manousos}} \yr{2010}  \at{Impulse and particle dislodgement
  under turbulent flow conditions}.  \jt{Physics of Fluids}  \bvol{22}~(4),
  \pg{046601}.

\bibitem[Diplas {\em et~al.\/}(2008)Diplas, Dancey, Celik, Valyrakis, Greer \&
  Akar]{diplas2008}
{\sc \au{Diplas, Panayiotis}, \au{Dancey, Clint~L}, \au{Celik, Ahmet~O},
  \au{Valyrakis, Manousos}, \au{Greer, Krista} \& \au{Akar, Tanju}} \yr{2008}
  \at{The role of impulse on the initiation of particle movement under
  turbulent flow conditions}.  \jt{Science}  \bvol{322}~(5902),  \pg{717--720}.

\bibitem[Fenton \& Abbott(1977)]{fenton1977}
{\sc \au{Fenton, JD} \& \au{Abbott, JE}} \yr{1977}  \at{Initial movement of
  grains on a stream bed: The effect of relative protrusion}.  \jt{Proc. R.
  Soc. Lond. A}  \bvol{352}~(1671),  \pg{523--537}.

\bibitem[Heathershaw \& Thorne(1985)]{heathershaw1985}
{\sc \au{Heathershaw, AD} \& \au{Thorne, PD}} \yr{1985}  \at{Sea-bed noises
  reveal role of turbulent bursting phenomenon in sediment transport by tidal
  currents}.  \jt{Nature}  \bvol{316}~(6026),  \pg{339}.

\bibitem[Kudrolli {\em et~al.\/}(2016)Kudrolli, Scheff \& Allen]{kudrolli2016}
{\sc \au{Kudrolli, Arshad}, \au{Scheff, David} \& \au{Allen, Benjamin}}
  \yr{2016}  \at{Critical shear rate and torque stability condition for a
  particle resting on a surface in a fluid flow}.  \jt{Journal of Fluid
  Mechanics}  \bvol{808},  \pg{397--409}.

\bibitem[Nelson {\em et~al.\/}(1995)Nelson, Shreve, McLean \&
  Drake]{nelson1995}
{\sc \au{Nelson, Jonathan~M}, \au{Shreve, Ronald~L}, \au{McLean, Stephen~R} \&
  \au{Drake, Thomas~G}} \yr{1995}  \at{Role of near-bed turbulence structure in
  bed load transport and bed form mechanics}.  \jt{Water resources research}
  \bvol{31}~(8),  \pg{2071--2086}.

\bibitem[Pantaleone \& Messer(2011)]{pantaleone2011}
{\sc \au{Pantaleone, J} \& \au{Messer, J}} \yr{2011}  \at{The added mass of a
  spherical projectile}.  \jt{American Journal of Physics}  \bvol{79}~(12),
  \pg{1202--1210}.

\bibitem[Schmeeckle {\em et~al.\/}(2007)Schmeeckle, Nelson \&
  Shreve]{schmeeckle2007}
{\sc \au{Schmeeckle, Mark~W}, \au{Nelson, Jonathan~M} \& \au{Shreve, Ronald~L}}
  \yr{2007}  \at{Forces on stationary particles in near-bed turbulent flows}.
  \jt{Journal of Geophysical Research: Earth Surface}  \bvol{112}~(F2).

\bibitem[Sumer {\em et~al.\/}(2003)Sumer, Chua, Cheng \& Freds{\o}e]{sumer2003}
{\sc \au{Sumer, B~Mutlu}, \au{Chua, Lloyd~HC}, \au{Cheng, N-S} \&
  \au{Freds{\o}e, J{\o}rgen}} \yr{2003}  \at{Influence of turbulence on bed
  load sediment transport}.  \jt{Journal of Hydraulic Engineering}
  \bvol{129}~(8),  \pg{585--596}.

\bibitem[Valyrakis {\em et~al.\/}(2013)Valyrakis, Diplas \&
  Dancey]{valyrakis2013}
{\sc \au{Valyrakis, Manousos}, \au{Diplas, Panayiotis} \& \au{Dancey, Clint~L}}
  \yr{2013}  \at{Entrainment of coarse particles in turbulent flows: An energy
  approach}.  \jt{Journal of Geophysical Research: Earth Surface}
  \bvol{118}~(1),  \pg{42--53}.

\bibitem[Valyrakis {\em et~al.\/}(2010)Valyrakis, Diplas, Dancey, Greer \&
  Celik]{valyrakis2010}
{\sc \au{Valyrakis, Manousos}, \au{Diplas, Panayiotis}, \au{Dancey, Clint~L},
  \au{Greer, Krista} \& \au{Celik, Ahmet~O}} \yr{2010}  \at{Role of
  instantaneous force magnitude and duration on particle entrainment}.
  \jt{Journal of Geophysical Research: Earth Surface}  \bvol{115}~(F2).

\end{thebibliography}

\end{document}